\documentclass[draft]{agujournal2019}
\justifying
\usepackage{url} 
\usepackage{lineno}
\usepackage[inline]{trackchanges} 
\usepackage{soul}
\usepackage{apacite}
\usepackage{amssymb}
\usepackage{gensymb}
\usepackage{float}
\usepackage{caption}
\usepackage{subcaption}
\usepackage{booktabs} 
\usepackage{threeparttable} 

\draftfalse

\journalname{JGR: Oceans}

\begin{document}

%
%


\title{Distribution of plastics of various sizes and densities in the global ocean from a 3D Eulerian model}

%
%




\authors{Zih-En Tseng\affil{1},
 Yue Wu\affil{1},
 Dimitris Menemenlis\affil{2},
 Guangyao Wang\affil{1},
 Chris Ruf\affil{3},
 Yulin Pan\affil{1}}

\affiliation{1}{Naval Architecture and Marine Engineering, the University of Michigan, MI, United States}
\affiliation{2}{Jet Propulsion Laboratory, California Institute of Technology, CA, United States}
\affiliation{3}{Climate and Space Sciences and Engineering, the University of Michigan, MI, United States}






\correspondingauthor{Yulin Pan}{yulinpan@umich.edu}



\begin{keypoints}
\item We develop a 3D Eulerian model to study the transport of microplastics in the global ocean, taking into consideration the effect of size and density of particles to their vertical terminal velocity. 

\item Stationary patterns from particles of different properties are studied, among which only low-density particles with sufficient size ($\gtrsim 10 \mu m$) accumulate in subtropical gyres observed in previous studies.

\item Seasonal variation of the surface particle concentration is demonstrated, which reasonably agrees with satellite observation by CYGNSS and is physically attributed to the seasonal variation of ocean mixing layer (ML) depth.
\end{keypoints}

%
%

%
%


\begin{abstract}
We develop a 3D Eulerian model to study the transport and distribution of microplastics in the global ocean. Among other benefits that will be discussed in the paper, one unique feature of our model is that it takes into consideration the effect of properties of particles (size and density, the former for the first time) to their vertical terminal velocity. With ocean current velocity taken from ECCOv4r4, a dataset generated from a data-assimilated MITgcm reanalysis, our model is integrated for 26 years for particles of different properties with their stationary patterns studied. We find that only low-density particles with sufficient size (e.g. density $900kg/m^3$ with size $\gtrsim 10 \mu m$) aggregate in the five subtropical gyres observed in previous studies. In contrast, particles of smaller size ($\sim 1 \mu m$), irrespective of their density, behave like neutrally buoyant particles with a weaker pattern on the surface and a deeper penetration into depth (up to about 1km deep). In addition, we observe seasonal variations of floating particle concentration on the ocean surface, which reasonably agree with the satellite observation by Cyclone Global Navigation Satellite System (CYGNSS) in terms of the phase of the variation. We find that the seasonal variation of the surface particle concentration correlates well with the variation of the mixing layer (ML) depth globally, due to an almost uniform vertical distribution of particles in the ML with total amount of particles conserved.
\end{abstract}

\section*{Plain Language Summary}
Oceanic plastic pollution is an urgent and global problem. We study the transport and distribution of microplastics (plastic pieces with sizes smaller than 5mm) through a newly developed computer simulation. In our simulations, the effect of plastic particle size is for the first time considered which affects the vertical motion of the particles together with their density. This consideration leads to new understanding of the microplastic stationary pattern compared to previous studies. For example, only plastic particles made from low-density material with sufficient size ($\gtrsim 10 \mu m$) accumulate in the five subtropical gyres forming the garbage patch identified in previous studies. Small enough ($\sim 1 \mu m$) plastic particles, on the other hand, exhibit weaker surface patterns and can be found in deeper ocean (up to about 1km deep). Additionally, we find that the amount of floating plastics on the sea surface changes seasonally, generally with more in summer and less in winter. This seasonal pattern agrees with the satellite observation by Cyclone Global Navigation Satellite System (CYGNSS), which consists of eight micro-satellites to measure, for example, wind speeds over Earth's oceans. We finally explain the seasonal variation and physically connect it to the variation of mixing layer depth of the ocean.

\section{Introduction}
Plastic pollution in the ocean is a fundamental environmental problem we face now and will face in the foreseeable future. An estimated eight million tons of plastic trash enters the ocean each year, and most of it is battered by sun and ocean waves into microplastics (i.e. plastic particles less than 5mm long according to a definition by NOAA on \url{https://marinedebris.noaa.gov/what-marine-debris/microplastics}). These micro pieces of plastics can easily enter the food chain and pollute the food we humans consume. Therefore, it is crucial to understand the distribution of microplastics in the global ocean for strategic removal of their pollution.

A major method for field measurements of oceanic plastic concentration is through trawler nets, which captures particles of size greater than 0.2mm floating near the ocean surface \cite{C_zar_2014,Eriksen_2014}. In spite of several successful attempts made using this approach, the information gained is too scarce in space and time. Moreover, the short length/time scale of the measurement also tends to make the results sensitive to inhomogeneous disturbances such as eddies and surface wind \cite{Lebreton_2012}. Remote sensing methods, on the other hand, provide much greater spatial coverage. Among several proposed techniques, \citeA{Maddy_Chris_2022} retrieves the concentration of floating microplastics from the surface roughness anomaly (i.e. surface roughness lower than expected from wind), with the first microplastics database covering $38\degree S - 38\degree N$ established. While favorable agreement with other approaches have been reported, it should be noted that the retrieval algorithm relies on the presence of surfactant, instead of microplastics, to damp the surface roughness \cite{Sun_2023}. Therefore, the results obtained need to be interpreted with caution on the difference of transport mechanisms between microplastics and surfactant. 

Transport models of microplastics offer another way to understand their global distribution. Table \ref{tbl:models_compare} summarizes the existing global transport models, classified as 2D Lagrangian, 2D statistical and 3D Eulerian models. In 2D Lagrangian models, each particle released to the ocean is tracked as a passive scalar following the ocean surface current velocity, obtained from global reanalysis datasets \cite{Chenillat_2021,Lebreton_2012}. In 2D statistical models, the surface microplastics concentration is evolved taking into consideration the advection velocity in a probabilistic way, with the probability calculated from the Global Drifter Program \cite{Niiler_2001,van_Sebille_2012,Maximenko_2012}. \citeA{Mountford_2019} represents the first 3D Eulerian model which solves the advection-diffusion particle equation in the global ocean. The vertical movement of particles is also for the first time incorporated, which depends on both vertical convective motion of ocean flow and the buoyancy effect resulted from the density difference between particles and sea water. When these models are integrated for sufficient time, it is usually found that accumulation of particles, or low-density particles in \citeA{Mountford_2019}, in five subtropical gyres appear as stationary patterns in the global ocean. However, it is not clear how this result is affected by particle size, which can affect the vertical motion of particles but is not considered in \citeA{Mountford_2019}. In addition, some model results reveal seasonal variations of the surface particles \cite{van_Sebille_2012}, which is also observed in remote sensing data \cite{Maddy_Chris_2022}, but a physical explanation of the phenomenon is lacking.

In this paper, we continue with the route of using Eulerian models to understand microplastics concentration in the global ocean. Indeed, not only have Eulerian models been extended to a 3D field, but also they have other benefits compared to Lagrangian models, e.g., the independence of computational cost on number of particles and extendable nature to incorporate particle fragmentation and data assimilation (as future work). Compared to \citeA{Mountford_2019}, the Eulerian model developed for this paper is also associated with several advantages: (1) we formulate the buoyancy advection term in a mass-conserved manner, together with a formula for the vertical terminal velocity which incorporates the effect of particle size; (2) Our ocean current field is sourced from the Estimating the Circulation and Climate of the ocean Version 4, Release 4 (ECCOv4r4) dataset. In particular, ECCOv4r4 is established from a MITgcm (MIT general circulation model) global simulation from 1992 to 2017, with constrained data from Jason-3 satellite mission and the Ice-Tethered-Profiler (ITP) mission, just to mention a few. As a result, the velocity data exhibits higher fidelity (e.g., Stokes drift by surface waves partially captured because of the constraint by surface drifter data) and interannual variability, with both features not incorporated in \citeA{Mountford_2019}. In addition, ECCOv4r4 is provided with a horizontal resolution of $40-100km$ and 50 vertical layers, and all our simulations are conducted at the same resolution which is almost two times finer in each spatial dimension than that in \citeA{Mountford_2019} (see Table \ref{tbl:models_compare}).

Equipped with our new Eulerian model, we study the stationary distribution pattern of microplastics with different properties (i.e., density and size) after integration of the model from 1992 to 2017. We find that only positively buoyant particles with sufficient size (e.g., density $\rho=900kg/m^3$ and diameter $d \gtrsim 10 \mu m$) form patches in the five subtropical gyres observed in previous work. For small enough particles ($d \lesssim 1 \mu m$) irrespective of their density, the distribution converges to that of neutrally buoyant particles with a much weaker surface accumulation pattern and a penetration into greater depth of about 1km. Additionally, for floating particles, we observe a seasonal variation of the surface concentration that reasonably agrees with remote sensing data from CYGNSS in terms of phase of the cycle. The seasonal variation of surface concentration is found to be well correlated with the seasonal variation of the mixing layer (ML) depth globally, based on which a physical explanation is developed. Specifically, as demonstrated in our model results, the floating particles have a nearly uniform distribution in the mixing layer due to the diffusion effect, and diminishes sharply at deeper depth. Because of the constant total mass of particles, their surface concentration varies inversely proportional to the depth of the ML (i.e., a column of uniformly distributed particles reacting to the variation of the ML depth).

This paper is organized as follows, in $\S$\ref{sec:methodology}, we provide details of our microplastics transport model, including the model equations and formulation to incorporate the vertical motion and terminal velocity, as well as details of the ECCOv4r4 dataset. In $\S$\ref{sec:results}, we show results of stationary patterns of plastic particles with different densities and sizes, with a focus on the effect of size to the distribution. The seasonal variation of the surface particles are discussed, in terms of its correlation with CYNGSS observation and the associated physical mechanism. $\S$\ref{sec:conclusion} concludes the work and points out possible directions for further improvements.


\begin{table}
 \begin{threeparttable}[b]
 \caption[\TeX{} engine features]{Summary of features of our model and previous ones.}
 \label{tbl:models_compare}
 \centering
 \begin{tabular}{r c c l}
 \midrule 
 Models & Velocity data source & Resolution & Model type \\
 \midrule
 Our model & ECCOv4r4\footnote{} & $1/2\degree-1\degree$ & 3D Eulerian \\
 \citeA{Chenillat_2021} & GLORYS\footnote{} & $1/12\degree$ & 2D Lagrangian \\
 Mountford $\&$ \\ Morales Maqueda (2019) & ORCA2-LIM3\footnote{} & $1\degree-2\degree$ & 3D Eulerian \\
 \citeA{Lebreton_2012} & HYCOM\footnote{} & $1/12\degree$ & 2D Lagrangian \\
 \citeA{van_Sebille_2012} & GDP\footnote{} & $1\degree$ & 2D statistical \\
 \citeA{Maximenko_2012} & GDP & $1\degree$ & 2D statistical \\
 \midrule 
 \end{tabular}
 \begin{tablenotes}
 \item [1] Estimating the Circulation and Climate of the Ocean Version 4 Release 4.
 \item [2] Global Ocean Physics Reanalysis.
 \item [3] Nucleus for European Modelling of the Ocean Version 3.6, configuration ORCA2‐LIM3.
 \item [4] HYbrid Coordinate Ocean Model.
 \item [5] Global Drifter Program.
 \end{tablenotes}
 \end{threeparttable}
\end{table}

\section{Methodology}
\label{sec:methodology}

We model the microplastic particles as passive particles advected by the ocean current. Mathematically, the concentration of particles is described by an advection-diffusion equation with an additional term to represent the buoyancy effect:

\begin{linenomath*}
\begin{equation}
\frac{\partial \tau}{\partial t} 
+ \nabla \cdot (\tau \mathbf{u})
+ \partial_z (\tau w_r)
= 
\nabla \cdot (K \nabla \tau)
+ Q,
\label{eq:adv-diff}
\end{equation}
\end{linenomath*}
where $\tau$ is the particle concentration in $g/m^3$, $\nabla \equiv \frac{\partial}{\partial x} \hat{i} + \frac{\partial}{\partial y} \hat{j} + \frac{\partial}{\partial z} \hat{k}$ is the 3D gradient operator, with $x$, $y$, $z$ the three spatial dimensions and $\hat{i}$, $\hat{j}$, $\hat{k}$ the unit vector in each dimension respectively, $\mathbf{u}$ is the ocean current velocity taken from the dataset ECCOv4r4 (described in detail in \S \ref{sec:eccov4r4}). The term $\partial_z (\tau w_r)$ represents the buoyancy effect with $w_r$, the vertical terminal velocity (described in \S \ref{sec:w_term}). $K$ is the $3\times3$ eddy diffusivity tensor, constructed by the summation of Gaspar-Grégoris-Lefevre (GGL) \cite{Gaspar_1990} and Gent-McWilliams eddy $\&$ Redi (GMRedi) \cite{Gent_1990} parameterizations, the former modeling diapycnal diffusivity and latter accounting for corrections due to tilted isopycnals. In general, the two mixing schemes depend on the turbulent kinetic energy and sea water density distribution respectively, with both globally available from the ECCOv4r4 dataset. Finally, $Q$ is the source term distributed along coastlines based on Jambeck’s global mismanaged waste survey, consistent with the coastal input used in \citeA{van_Sebille_2012} and \citeA{Mountford_2019}.

In this work, equation (\ref{eq:adv-diff}) is solved using the `ptracers' package in MITgcm, with necessary modifications to include the additional terms. Numerical setups for the simulation are summarized in $\S$\ref{sec:num_setup}, with key simulation files, result animations, and procedures made available and summarized in the section ``Open Research Section''. Animations of some simulation results are also uploaded as supplemental files.


\subsection{ECCOv4r4 dataset}
\label{sec:eccov4r4}
Estimating the Circulation and Climate of the Ocean Version 4 Release 4 (ECCOv4r4) provides a data-constrained ocean state reanalysis from 1992 to 2007 \cite{ecco_v4r4_synopsis,eccov4r4_data,Forget_2015}. The backbone of this reanalysis is an MITgcm global simulation, together with a 4D-var technique to assimilate measurement data from sources such as Argo and OCCA (Ocean Comprehensible Atlas) \cite{Toole_2011,Forget_2010}. The MITgcm simulation is conducted in hydrostatic mode on the so-called ECCO-LLC90 grid, which has about $1\degree$ horizontal resolution (corresponding to $40-100km$ grid size) with further meridional refinement and $50$ vertical layers with $10-500m$ resolution. A time step of $1hr$ is used which ensures the stability of the simulation. Parameters such as initial conditions and forcing (biweekly averaged surface wind stress, radiative heat flux, etc.) are set as variables, through the tuning of which the difference between the simulation and measurements are minimized as in the 4D-var technique. 

We take the ocean current velocity $\mathbf{u}$ from ECCOv4r4 to feed the simulation of equation (\ref{eq:adv-diff}). Such a velocity field contains resolved physics such as geostrophic and the Ekman current components, as well as other components captured in an indirect way. For example, surface wave induced Stokes drift is partially captured due to the adjustment of wind shear stress to fit the data from surface drifter as part of the data assimilation \cite{Reynolds_2002}. In addition, MITgcm simulation also includes an equation for turbulent kinetic energy (with parameterized production and dissipation) which provides information to calculate the GGL parameterization of diffusivity in equation (\ref{eq:adv-diff}) and for temperature and salinity. 

\subsection{Buoyancy advection term}
\label{sec:w_term}
The buoyancy effect of particles is modeled by a term in advection form, $\partial_z(\tau w_r)$, in equation (\ref{eq:adv-diff}). This is because a particle falling/rising with its terminal velocity $w_r$ is equivalent to the particle being advected by an ``external'' velocity $w_r$. We note that this buoyancy advection term has to be in a mass-conservative form, in the sense that if there is no flux (or balanced flux) from the upper and lower boundary the mass in the domain should stay constant. This property is satisfied only when the term is written as $\partial_z(\tau w_r)$ instead of $w_r\partial_z\tau $. In \citeA{Mountford_2019}, the latter form is used at least judged from their presented equation, which may result in inconsistency in some presented results. 

We now discuss the formula of $w_r$ dependent on both the particle size and density, as presented in \citeA{Yang_2014} and \citeA{Dey_2019}. Consider a spherical particle of diameter $d$ at terminal velocity $w_r$ subject to buoyancy force $B= g (\rho_w-\rho_p) (4\pi/3) (d/2)^3$ and drag force $D=C_D (1/2) (\rho_w w_r^2) \pi (d/2)^2$, with their balance leading to
\begin{linenomath*}
\begin{equation}
w_r^2 C_D = \frac{4}{3} dg \frac{\rho_w-\rho_p}{\rho_w},
\label{eq:wr_balance}
\end{equation}
\end{linenomath*}
where $g$ is the gravitational acceleration, $\rho_w$ is the water density, $\rho_p$ is the particle's mass density, $C_D$ is the drag coefficient that depends on the particle Reynolds number $Re \equiv \rho_w d w_r / \mu$, with $\mu$ the dynamic viscosity of sea water. For typical size ($ < 0.1 mm$) and velocity ($ < 1 mm/s$) of microplastics, we have $Re=O(0.1)<1$. At this regime of $Re$, it has been shown that $C_D=24/Re$ irrespective of the shape of the particle. Therefore, the final formula can be written as
\begin{linenomath*}
\begin{equation}
w_r = \frac{g(\rho_w-\rho_p)d^2}{18\mu},
\label{eq:wr_stokes}
\end{equation}
\end{linenomath*}
which we use in the simulation of equation (\ref{eq:adv-diff}). As an example, consider biomaterial particles like phytodetritus, typically at mass density $1388kg/m^3$ \cite{Kooi_2017}. The formula (\ref{eq:wr_stokes}) gives sinking velocity $42.7m/day$ and $170.9m/day$ for particles at $50\mu m$ and $100\mu m$ respectively, which is comparable to the measurement of at least $64m/day$ in the marine snow event after the 2011 Tohoku–Oki Earthquake (although the size of particles in the measurement is not specified) \cite{Oguri_2013}. In addition, based on the buoyancy and drag forces to accelerate the particles, the typical time for a microplastic particle to reach terminal velocity is below 1s (see \ref{sec:append_t_e_for_w_r} for a detailed analysis). As a result, the acceleration process can be safely neglected in our model with the total integration time of 26 years. 

\subsection{Numerical and Case Setup}
\label{sec:num_setup}
\begin{figure}[h!]
\centering
\noindent\includegraphics[width=.7\linewidth]{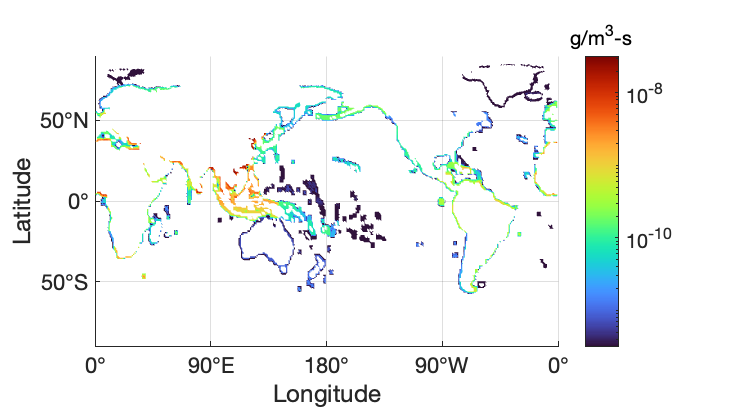}
\caption{Map of $Q$ on the ECCO-LLC90 grid, based on the waste management report \protect\cite{Jambeck_2015}.}
\label{fig:source}
\end{figure}

We numerically integrate equation (\ref{eq:adv-diff}) with a finite volume method using the MITgcm ``ptracers'' package with additional implementation of the buoyancy advection term. The computation is performed on the ECCO-LLC90 grid with the same resolution and time step as those to generate the ECCOv4r4 dataset. At all boundaries, zero flux boundary condition (including the flux due to $w_r$) is applied. Since MITgcm simulation is conducted with a variable free surface (instead of a rigid lid), the thickness of cells need to be changed at each time step to accommodate the variation of free surface. This is treated through a mapping technique presented in \citeA{Adcroft_2004,Campin_2004}. The $Q$ term in equation (\ref{eq:adv-diff}) is handled by the ``gchem'' package, with positive values in $g/m^3s$ converted from the mismanaged weight and distributed in the surface cells along the coastlines. In the situation where a cell is cut by the coastline, a ``partial-step'' method is applied to account for the situation \cite{Adcroft_1997}. A map of $Q$ on the ECCO-LLC90 grid is provided in figure \ref{fig:source}.

We consider two idealistic cases and six realistic cases as summarized in Table \ref{tbl:experiments}, with the former providing insight on the particle behavior under extreme conditions and the latter aiming to understand the effect of particle properties on the distribution. The first idealistic case is when the problem is reduced to two dimensions (2D), which is an extreme case for particles with sufficiently high buoyancy ($w_r \to \infty$) so that they stay on the top layer near the free surface. This case is practically simulated by setting the third component of both $\mathbf{u}$ and gradient operator, as well as $w_r$, to be zero in equation (\ref{eq:adv-diff}). It can be hypothesized that the 2D idealistic case leads to strong gyre patterns since particle accumulation in converging flow zone in the ocean cannot be reduced by the corresponding vertical downward flow as in 3D cases. For an analytical solution of 2D accumulation, see \citeA{Maximenko_2012}. The second idealist case is for neutrally buoyant particles to represent those with $d\rightarrow0$ so that $w_r\rightarrow0$. The six realistic cases are conducted for three types of plastics (PE for Polyethylene, PP for Polypropylene, and PVC for Polyvinyl Chloride) with densities lower than, comparable to and higher than sea water, respectively. For each particle density, two particle sizes are considered for their distinguished behavior in the stationary distribution. 

\begin{table}
 \begin{threeparttable}[b]
 \caption[\TeX{} engine features]{Summary of particle properties in the two idealistic and six realistic cases. A typical value of terminal velocity is estimated based on a sea water density $1025kg/m^3$.}
 \label{tbl:experiments}
 \centering
 \begin{tabular}{l c c c}
 \hline
 Particle label & Mass density($kg/m^3$) & Diameter ($\mu m$) & $w_r(\rho_w=1025kg/m^3)$\\
 \hline
 2D surface & NA & NA & NA \\
 Neutrally buoyant & NA & NA & 0 \\
 PE-10 & 900 & 10 & $6.3\mu m/s$ \\
 PE-1 & 900 & 1 & $0.06\mu m/s$ \\
 PP-100 & 1030 & 100 & $-27.3\mu m/s$\\
 PP-10 & 1030 & 10 & $-0.27\mu m/s$\\
 PVC-10 & 1200 & 10 & $-9.5\mu m/s$ \\
 PVC-1 & 1200 & 1 & $-0.09\mu m/s$\\
 \hline
 \multicolumn{4}
 {l}{Note: typical vertical ocean current has magnitude around $5 \mu m/s$.}
 \end{tabular}
 \end{threeparttable}
\end{table}

\section{Results}
\label{sec:results}
\subsection{Idealistic cases}
\begin{figure}[h!]
\centering
\begin{subfigure}{\textwidth}
 \centering
 \includegraphics[width=.7\linewidth]{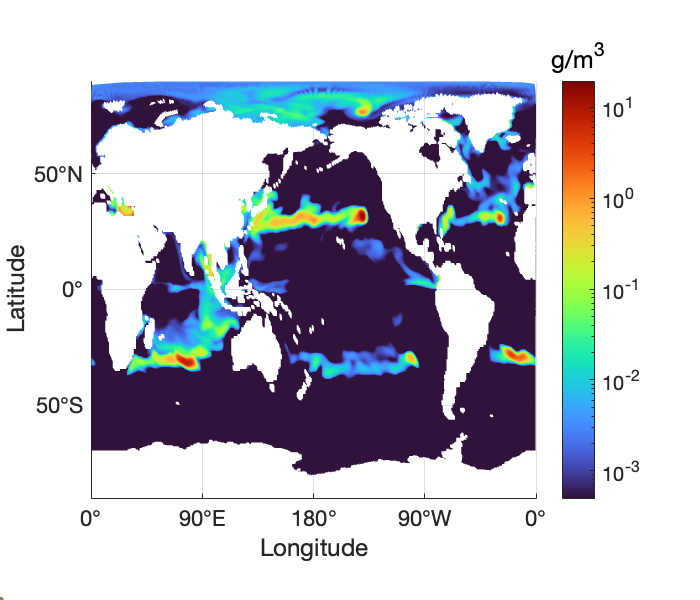}
\end{subfigure}
\caption{Surface particle concentration in the 2D idealistic case, averaged over the last year of the 26-year simulation.}
\label{fig:2D}
\end{figure}

We first present results for the 2D idealistic cases in figure \ref{fig:2D}, which shows the surface concentration of particles averaged over the last year of the 26-year simulation. Together with our results we also provide in \ref{append:graph_compare} a graphical comparison to results from previous 2D studies in \citeA{Maximenko_2012}, \citeA{Lebreton_2012}, \citeA{van_Sebille_2012}, and \citeA{Chenillat_2021}. It can be seen that our result reproduces accumulation in the five subtropical gyres at North Pacific, South Pacific, North Atlantic, South Atlantic, and Indian Ocean found in previous works. In addition, we observe another accumulation in the north Arctic area which is not present in aforementioned previous studies (for some of them because the ocean current data, say from GDP, does not well cover the arctic area). However, gyre accumulation patterns at the arctic have indeed been observed (although weaker) in the 3D Eulerian model for positively buoyant particles as shown in \citeA{Mountford_2019} (see their figure 3). In general, our results shown here is also a confirmation about our hypothesis in $\S$\ref{sec:num_setup} on the strong gyre accumulation pattern for the 2D case.

\begin{figure}[h!]
\centering
\begin{subfigure}{.6\textwidth}
 \centering
 \includegraphics[width=\linewidth]{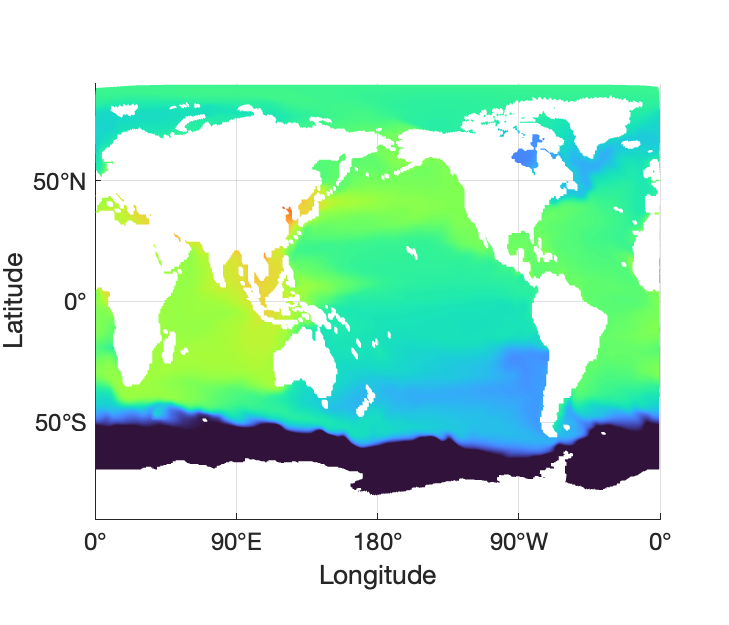}%
 \includegraphics[width=.14\linewidth]{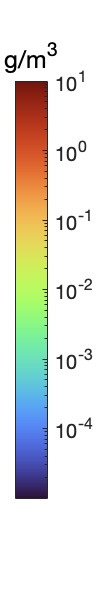}
 \caption{}
 \label{fig:neutral_z}
\end{subfigure}
\begin{subfigure}{.6\textwidth}
 \centering
 \includegraphics[width=\linewidth]{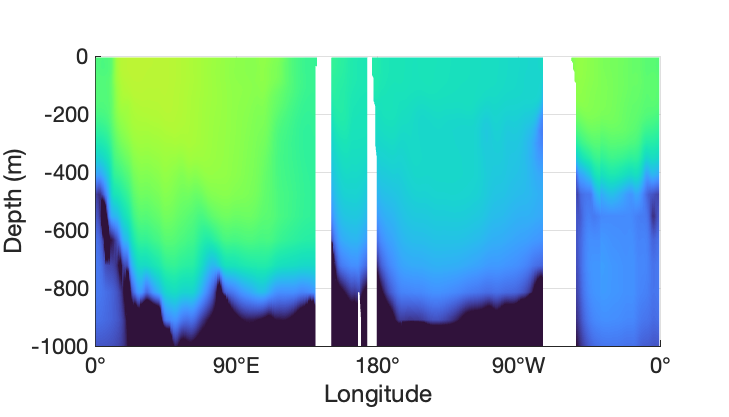}%
 \includegraphics[width=.14\linewidth]{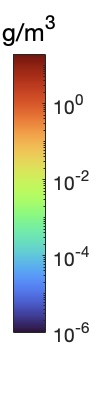}
 \caption{}
 \label{fig:neutral_y1k}
\end{subfigure}
\caption{Particle distribution in the case of neutrally buoyant particles, averaged over the last year of the 26-year simulation, in terms of concentration (a) on the surface and (b) over a vertical slice at $38\degree S$ and upper 1km of the ocean.}
\label{fig:neutral}
\end{figure}

We next show results for the case with neutrally buoyant particles in figure \ref{fig:neutral}, (a) for surface concentration and (b) for concentration at a vertical slice of $38\degree S$ up to $1km$ depth to reveal the vertical distribution. A graphical comparison of this result to that in \citeA{Mountford_2019} is also provided in Appendix B. It can be seen from figure \ref{fig:neutral_z} that the gyre accumulation pattern is much weaker in this case compared to the 2D result, with some weak accumulations present in North Pacific, North Atlantic, South Atlantic, and Indian Ocean, but not South Pacific. The vertical distribution in figure \ref{fig:neutral_y1k} shows that the particles penetrate downward up to about $1km$ depth, much deeper than the mixing layer depth of about $200m$. This penetration could be due to the diapycnal mixing present ubiquitously everywhere in the ocean interior. We will use the results in figure \ref{fig:neutral} as a reference to understand the distribution of realistic particles with sufficiently small size in following discussions. 

\subsection{Realistic cases}
\subsubsection{Positively buoyant particles}
\label{sec:PE}
\begin{figure}[h!]
\centering
\makebox[\textwidth][c]{%
    \begin{subfigure}{.5\textwidth}
        \centering
        \includegraphics[width=\linewidth]{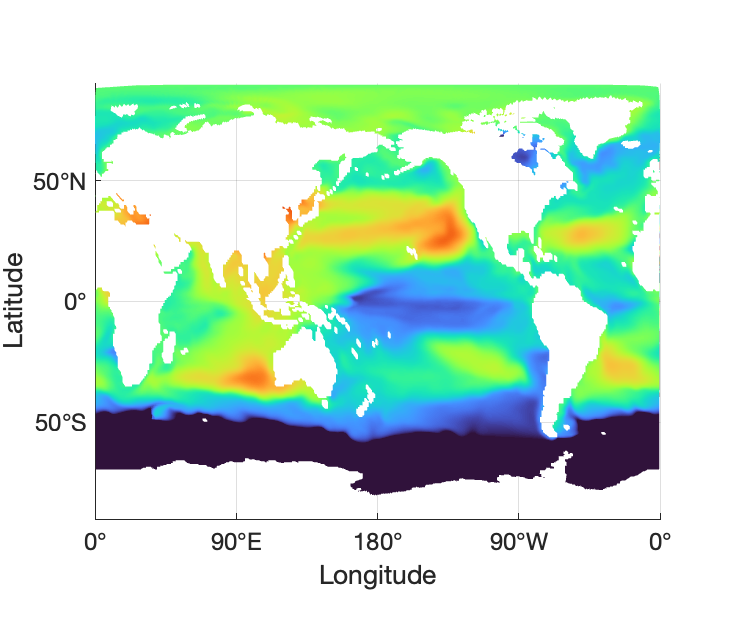}
        \caption{}
        \label{fig:PE-10_z}
        \end{subfigure}%
    \begin{subfigure}{.5\textwidth}
        \centering
        \includegraphics[width=\linewidth]{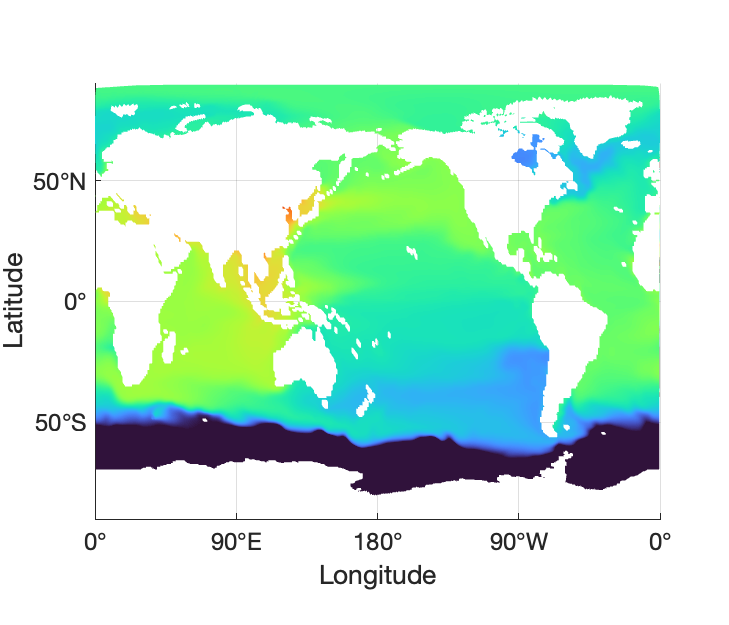}%
        \includegraphics[width=.14\linewidth]{figs/colorbar_z.png}
        \caption{}
        \label{fig:PE-1_z}
        \end{subfigure}
        }
\makebox[\textwidth][c]{%
    \begin{subfigure}{.5\textwidth}
        \centering
        \includegraphics[width=\linewidth]{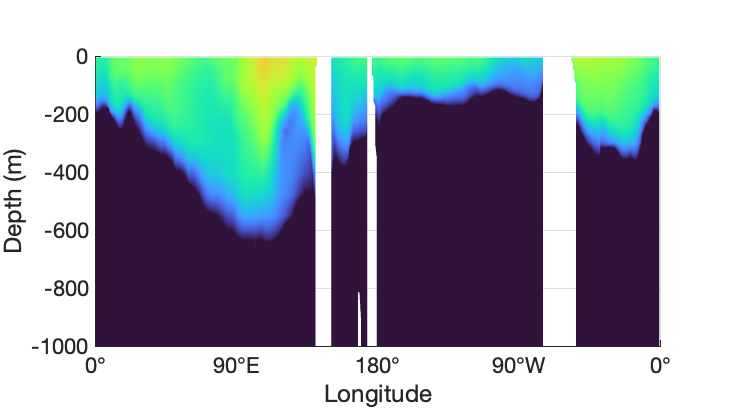}
        \caption{}
        \label{fig:PE-10_y1k}
        \end{subfigure}%
    \begin{subfigure}{.5\textwidth}
        \centering
        \includegraphics[width=\linewidth]{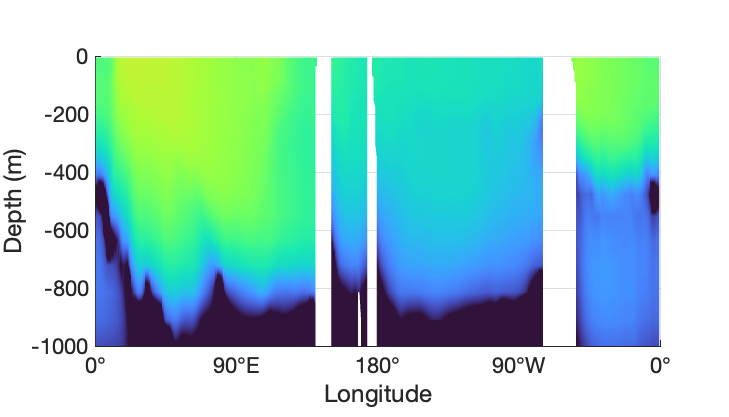}%
        \includegraphics[width=.14\linewidth]{figs/colorbar_y.png}
        \caption{}
        \label{fig:PE-1_y1k}
        \end{subfigure}
        }
\caption{Particle distribution of PE particles of two sizes, averaged over the last year of the 26-year simulation. The distribution of PE-10 is shown in terms of concentration (a) on the surface and (c) over a vertical slices at $38\degree S$, and the counterpart for PE-1 is shown in (b) and (d).}
\label{fig:PE}
\end{figure}

Polyethylene (PE) is a plastic material with density lower than the sea water. Here we take a representative density of  $900kg/m^3$, and two particle sizes of $10 \mu m$ (PE-10) and $1 \mu m$ (PE-1). Snapshots of the yearly-averaged solutions at the end of 26-year simulations for both particle sizes are shown in figures \ref{fig:PE}, in terms of surface concentration and vertical distributions. 

For PE-10 (i.e. larger particles), we see clear accumulation of particles in the five gyres identified in the 2D results, as well as some particle concentration in the Arctic ocean similar to figure \ref{fig:2D}. As expected, the concentration in the gyres is generally weaker than those shown in figure \ref{fig:2D} (note that different color bars are used) due to the vertical downwelling motion in the 3D cases. From figure \ref{fig:PE-10_y1k}, we find that the majority of particles stay close to the sea surface within 200m depth, i.e., the depth of ocean mixing layer. For PE-1 (i.e., smaller particles), in contrast, the accumulation in gyres becomes weaker and distributions shown in figures \ref{fig:PE-1_z} and \ref{fig:PE-1_y1k} are visually identical to the neutrally buoyant cases shown in figure \ref{fig:neutral}. A quantitative comparison between PE-1 and neutrally buoyant cases reveals some small difference in the concentration within the upper $500m$ of surface water. Indeed, for PE-1, the terminal velocity is $0.06\mu m/s$, which at most results in $50m$ upward shift of the particles in the 26 years of evolution.

Comparing results regarding PE-10 and PE-1, we find that larger particles are more likely to be accumulated in the gyres close to the surface. Therefore, ocean surface serves as a filter on the size of PE particles in the sense that larger particles are present in the gyres and smaller particles are ``sifted'' downward by the vertical mixing process. In the previous trawler experiments collecting the surface particles, it should be expected that most of particles with a similar density as PE are with a size above some threshold.

\subsubsection{Particles with density close to sea water}
\label{sec:PP}
\begin{figure}[h!]
\centering
\makebox[\textwidth][c]{%
    \begin{subfigure}{.5\textwidth}
        \centering
        \includegraphics[width=\linewidth]{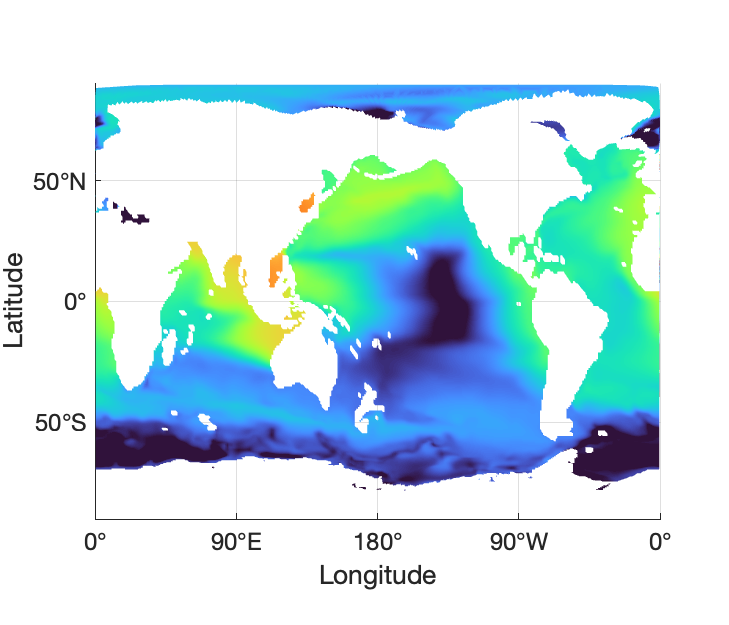}
        \caption{}
        \label{fig:PP-100_z}
        \end{subfigure}%
    \begin{subfigure}{.5\textwidth}
        \centering
        \includegraphics[width=\linewidth]{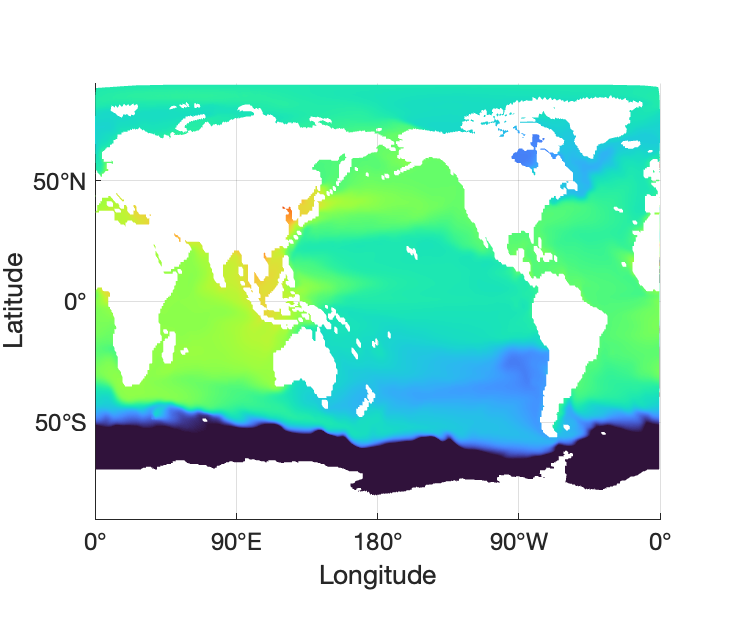}%
        \includegraphics[width=.14\linewidth]{figs/colorbar_z.png}
        \caption{}
        \label{fig:PP-10_z}
        \end{subfigure}
        }
\makebox[\textwidth][c]{%
    \begin{subfigure}{.5\textwidth}
        \centering
        \includegraphics[width=\linewidth]{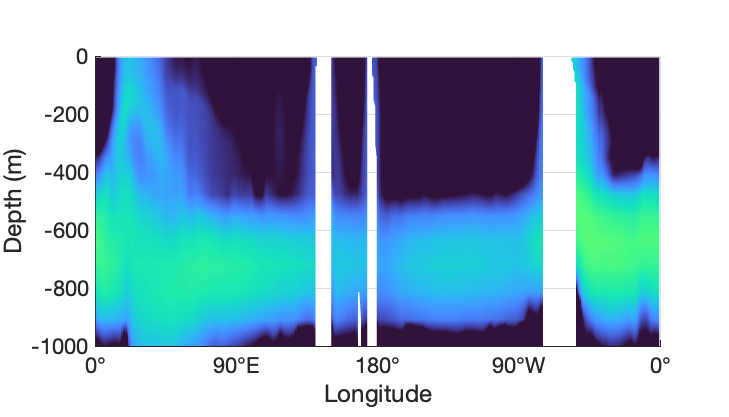}
        \caption{}
        \label{fig:PP-100_y1k}
        \end{subfigure}%
    \begin{subfigure}{.5\textwidth}
        \centering
        \includegraphics[width=\linewidth]{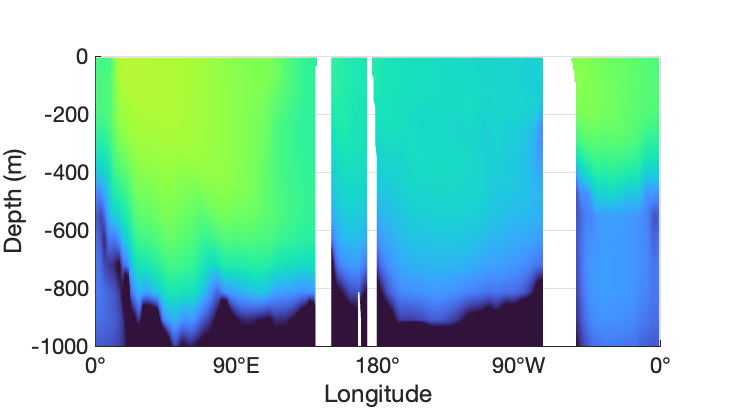}%
        \includegraphics[width=.14\linewidth]{figs/colorbar_y.png}
        \caption{}
        \label{fig:PP-10_y1k}
        \end{subfigure}
        }
\caption{Particle distribution of PP particles of two sizes, averaged over the last year of the 26-year simulation. The distribution of PP-100 is shown in terms of concentration (a) at $-650m$ depth and (c) over a vertical slices at $38\degree S$. The distribution of PP-10 is shown in terms of concentration (b) on the surface and (d) over a vertical slices at $38\degree S$.}
\label{fig:PP}
\end{figure}

Polypropylene (PP) is a plastic material with a density comparable to sea water. We take a representative density of $1030kg/m^3$, in two sizes of $10 \mu m$ (PP-10) and  $100 \mu m$ (PP-100), with the simulation results shown in figure \ref{fig:PP}. The PP-100 particles are mostly concentrated, as shown in figure \ref{fig:PP-100_y1k}, at a depth of about $650m$ where the local sea water density is close to particle density. This is a stable state for the stationary distribution since the PP-100 particles with any deviation from this depth tend to be returned to the depth by the buoyancy restoring effect (with a water density increasing with depth in the stably stratified state). At this depth of $650m$, the particle concentration is plotted in figure \ref{fig:PP-100_z}, which shows high concentration in regions close to the coast. This is because the particles released at the coastlines are advected horizontally to the open ocean and in the meanwhile sink vertically due to the buoyancy effect. As they reach about $650m$ depth, the horizontal advection almost vanishes due to small horizontal ocean flow motion at that depth. Therefore, there exists a time scale $T_s \approx 9 \ months$ determined by the time of particles sinking to the stationary layer, beyond which the particles are not further advected to the open ocean.

For the PP-10 results shown in figures \ref{fig:PP-10_z} and \ref{fig:PP-10_y1k}, in contrast, the distribution becomes similar to that in the neutrally buoyant case. We note that for density of $1030kg/m^3$, $10 \mu m$ is demonstrated to be the critical size for the neutrally buoyant behavior to occur, whereas for PE density of $900kg/m^3$ discussed in $\S$\ref{sec:PE}, the critical size goes down to $1 \mu m$. This can be understood from equation (\ref{eq:wr_stokes}), where $w_r$ can be made sufficiently small from either a small particle size $d$ or small density difference $\rho_w-\rho_p$. In addition, for the PP density, the ocean surface serves as a filter with an opposite effect to that for the PE density, in the sense that only small particles are expected to be found in trawler experiments at the ocean surface.  

\subsubsection{High-density particles}
\begin{figure}[h!]
\centering
\makebox[\textwidth][c]{%
    \begin{subfigure}{.5\textwidth}
        \includegraphics[width=\linewidth]{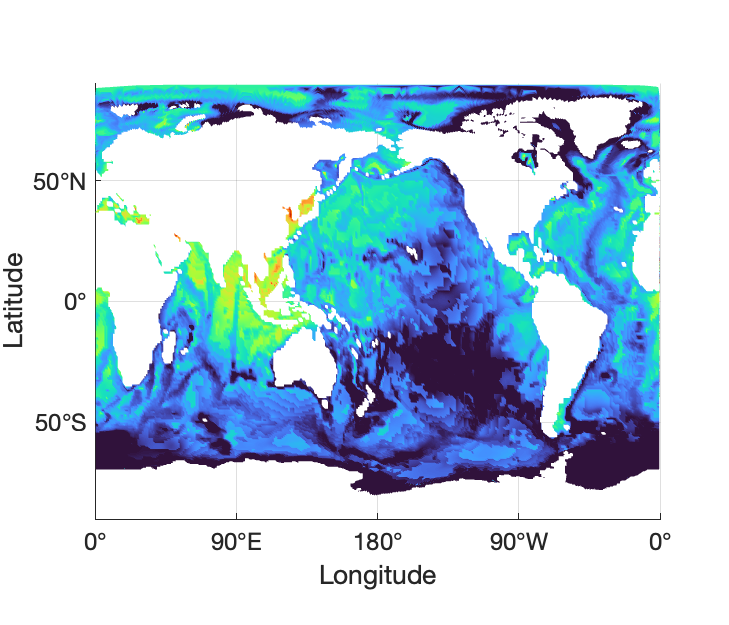}
        \caption{}
        \label{fig:PVC-10_z_bot}
        \end{subfigure}%
    \begin{subfigure}{.5\textwidth}
        \includegraphics[width=\linewidth]{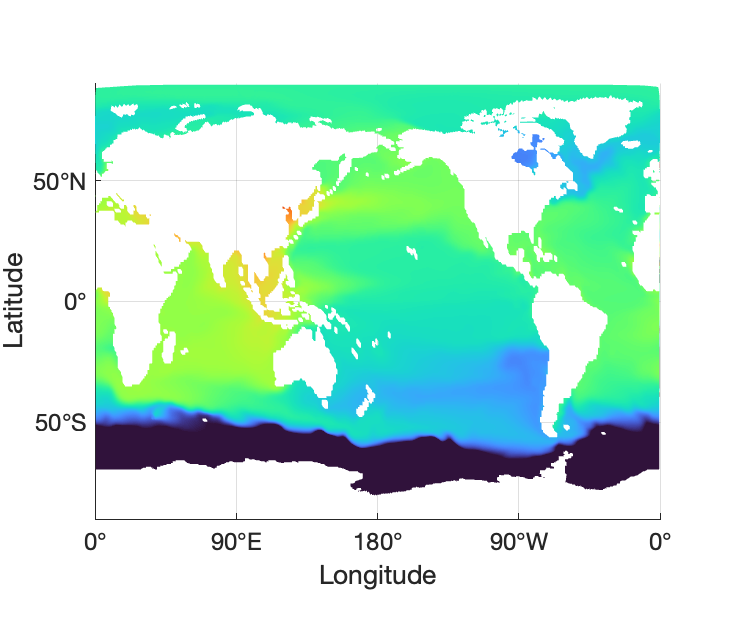}%
        \includegraphics[width=.14\linewidth]{figs/colorbar_z.png}
        \caption{}
        \label{fig:PVC-1_z}
        \end{subfigure}
        }
\makebox[\textwidth][c]{%
    \begin{subfigure}{.5\textwidth}
        \includegraphics[width=\linewidth]{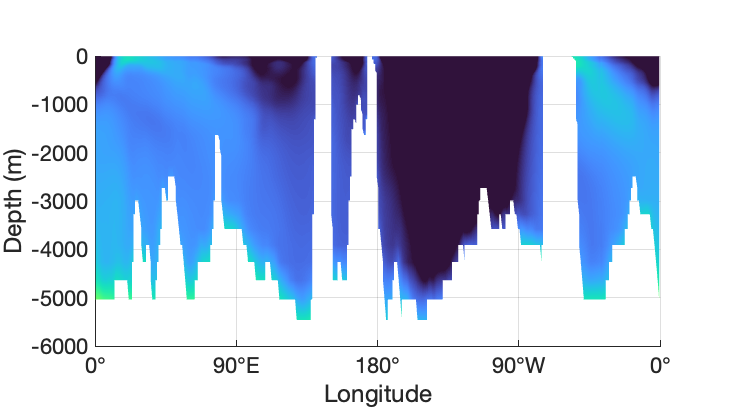}
        \caption{}
        \label{fig:PVC-10_y6k}
        \end{subfigure}%
    \begin{subfigure}{.5\textwidth}
        \includegraphics[width=\linewidth]{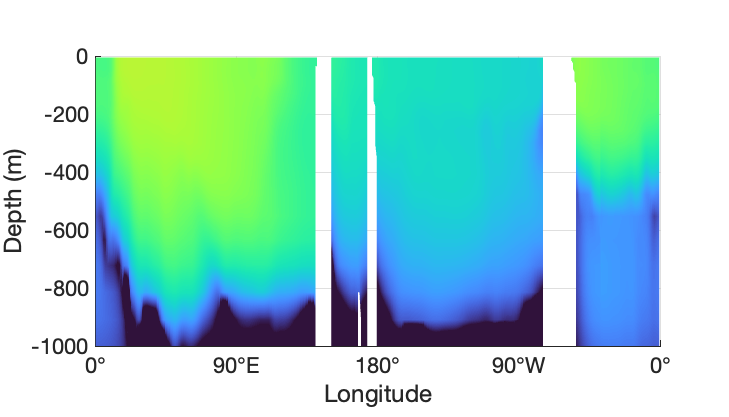}%
        \includegraphics[width=.14\linewidth]{figs/colorbar_y.png}
        \caption{}
        \label{fig:PVC-1_y1k}
        \end{subfigure}
        }
\caption{Particle distribution of PVC particles of two sizes, averaged over the last year of the 26-year simulation. The distribution of PVC-10 is shown in terms of concentration (a) at the ocean bottom and (c) over a vertical slices at $38\degree S$ covering all ocean depth. The distribution of PVC-1 is shown in terms of concentration (b) on the surface and (d) over a vertical slices at $38\degree S$ covering only the upper $1km$ depth.}
\label{fig:PVC}
\end{figure}

Polyvinyl chloride (PVC) is a plastic material with a density significantly higher than sea water. We take a representative density of $1200kg/m^3$ which is higher than the sea water density in the range of $1000<\rho_w<1050kg/m^3$. Figure \ref{fig:PVC} shows the results for two particle sizes $10 \mu m$ (PVC-10) and $1 \mu m$ (PVC-1). At the end of the 26-year simulation, most PVC-10 particles, as demonstrated in figure \ref{fig:PVC-10_y6k}, sink to the sea bottom and accumulate in the ocean trenches. However, there are some concentration of particles present over all depth as seen in figure \ref{fig:PVC-10_y6k}, which are from particles released from the coastline and in their process to sink to the bottom. We note that for PVC-10, it takes $T_s \approx 20 \ years$ for them to sink to the sea bottom with $w_r=-9.5 \mu m/s$. This value of $T_s$ is much longer than that for PP-100, so that particles over all depth in figure \ref{fig:PVC-10_y6k} is more evident than that in figure \ref{fig:PP-100_y1k} (since the PVC-10 particles have more time to travel horizontally before sinking to the bottom). Figure \ref{fig:PVC-10_z_bot} shows the sea bottom concentration of PVC-10 particles, which demonstrates higher concentration near the coastline similar to PP-100 in figure \ref{fig:PP}, but with a greater horizontal spreading for PVC-10 particles.

For PVC-1, as shown in figure \ref{fig:PVC-1_z} and \ref{fig:PVC-1_y1k}, the particle distribution again becomes similar to that of neutrally buoyant particles. A terminal vertical velocity of $w_r\approx 0.09 \mu m/s$ results in a downshift of $78 m$ in 26 years, which only slightly shifts down the concentration of PVC-1 compared to neutrally buoyant particles. As a result, the ocean surface serves as a filter in the same way as that for the PP particles.

\subsection{Seasonal variation of surface concentration}

\begin{figure}[h!]
\centering
\noindent\includegraphics[width=.7\textwidth]{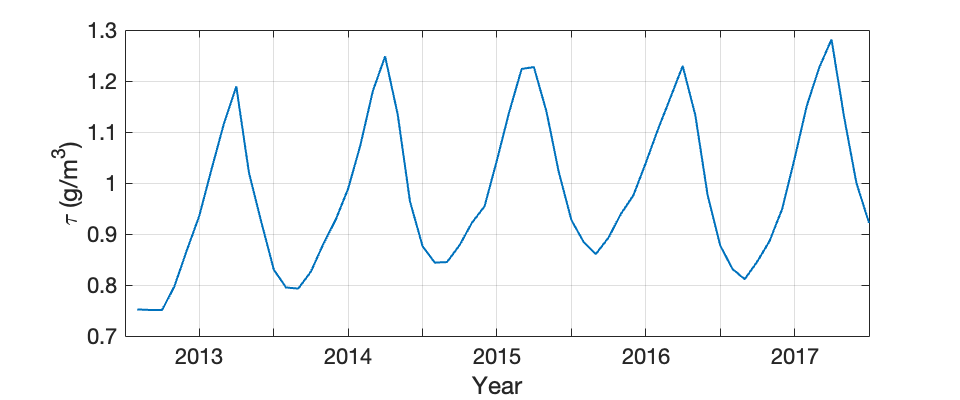}
\caption{Surface concentration of PE-10 microplastics averaged over the North Pacific [$20\degree N-35\degree N, 130\degree W-150\degree E$] across 5 years. In general, concentration is higher in summer and lower in winter.}
\label{fig:vs.CYGNSS:NPac}
\end{figure}

Seasonal variation of particle concentrations have been revealed in remote sensing data from CYGNSS \cite{Maddy_Chris_2022}, and later also recorded from the previous numerical study \cite{Mountford_2019}. In this section, we focus on the variation of surface concentration for which the most important plastic particles are those like PE-10. Figure \ref{fig:vs.CYGNSS:NPac} shows the variation of $\tau$ of PE-10 averaged over the North Pacific Ocean for 5 years, where we can see the variation within each year (i.e. seasonal variation) as well as variations over a longer time scale. It is the seasonal variation that we are interested in here.

\begin{figure}[h!]
\Centering
\makebox[\textwidth][c]{%
    \begin{subfigure}{.6\textwidth}
        \includegraphics[width=\linewidth]{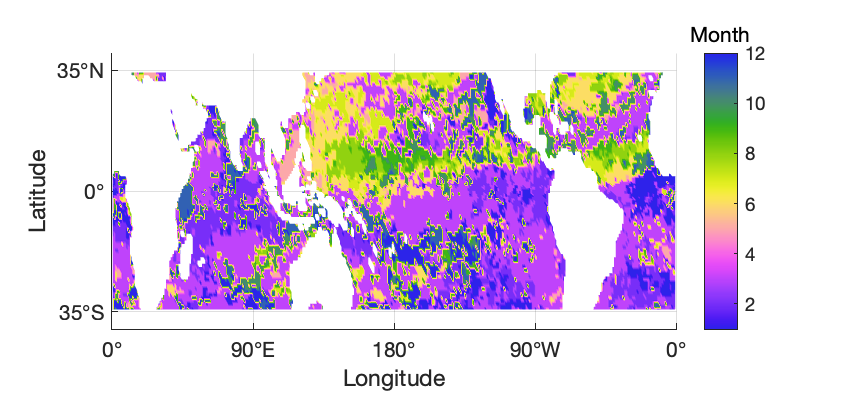}
        \caption{}
        \end{subfigure}%
    \begin{subfigure}{.52\textwidth}
        \includegraphics[width=\linewidth]{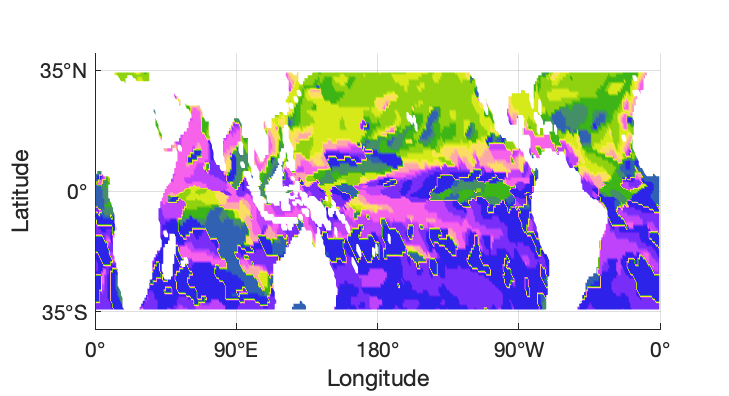}
        \caption{}
        \end{subfigure}
        }
\caption{Map of $T_p$ from (a) CYGNSS observation and (b) the PE-10 model result. The color denotes months 1-12, meaning January-December of a calendar year.}
\label{fig:vs.cygnss:Tp}
\end{figure}

We first perform a comparison between our model data and CYGNSS observation in terms of the phase of the seasonal cycle. For this purpose, we define the peak time, $T_p$, as the month within a year when maximum value of $\tau$ is recorded on the ocean surface. Figure \ref{fig:vs.cygnss:Tp} shows maps of $T_p$ between $38\degree S$ and $38\degree N$ (i.e., the covered region of CYGNSS measurement) in year 2017 from both our model data and CYGNSS data. In particular, the CYGNSS result is computed from the version 1.0 CYGNSS level 3 ocean microplastics concentration data at $1/4\degree$ resolution. We see that the comparison in figure \ref{fig:vs.cygnss:Tp} shows reasonable (first-order) agreement, e.g., in terms of peak concentration in summer in the subtropical bands of both the Northern and Southern hemispheres. Of course, certain level of discrepancy exists, which should be understood taking consideration of many factors. For example, CYGNSS retrieves the microplastic concentration from the surface roughness, which is dominantly affected by surfactant \cite{Sun_2023} (along with other factors such as precipitation and sea surface air pressure), which may have a different seasonal variation mechanism from plastic particles.  

\begin{figure}[h!]
\centering
\makebox[\textwidth][c]{%
    \begin{subfigure}{.6\textwidth}
        \centering
        \includegraphics[width=\linewidth]{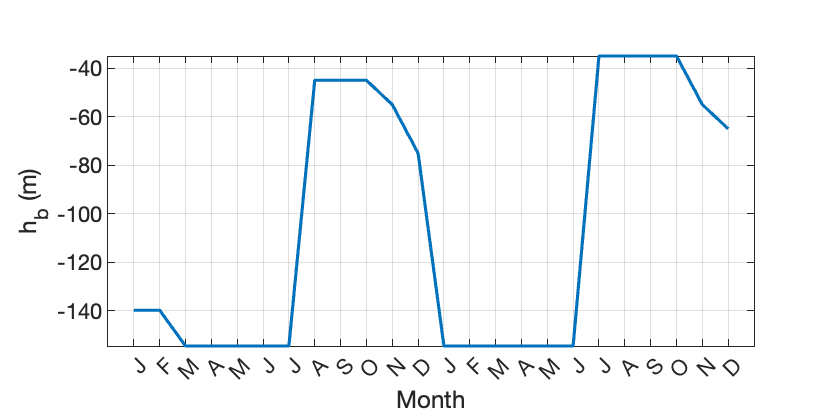}
        \caption{}
        \end{subfigure}%
    \begin{subfigure}{.6\textwidth}
        \centering
        \includegraphics[width=\linewidth]{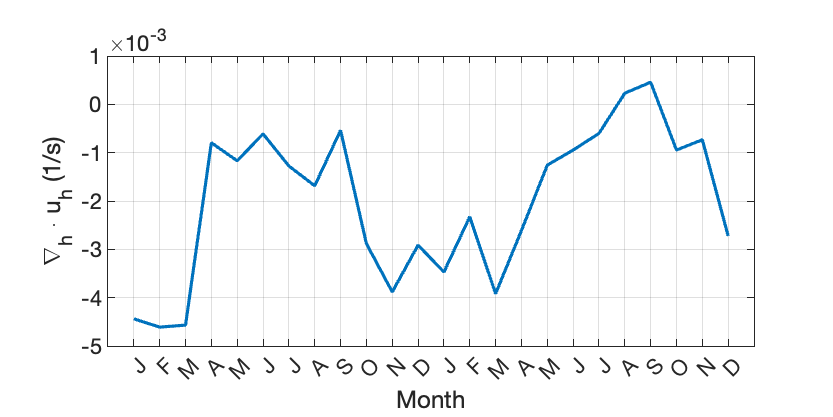}
        \caption{}
        \end{subfigure}
        }
\makebox[\textwidth][c]{%
    \begin{subfigure}{.6\textwidth}
        \centering
        \includegraphics[width=\linewidth]{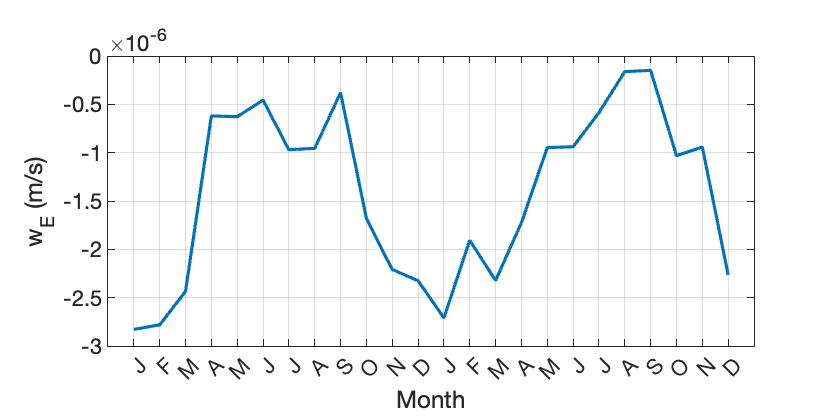}
        \caption{}
        \end{subfigure}%
    \begin{subfigure}{.6\textwidth}
        \centering
        \includegraphics[width=\linewidth]{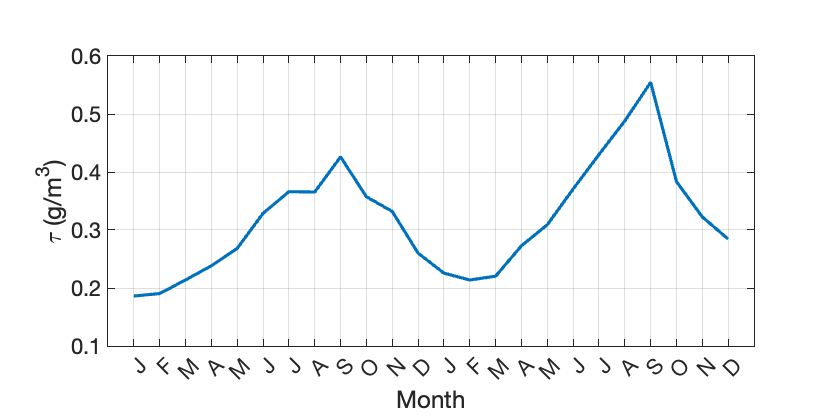}
        \caption{}
        \end{subfigure}
        }
\caption{Two years of seasonal cycle of (a) mixed layer depth $h_b$, (b) ocean current horizontal divergence $\nabla_h \cdot \mathbf{u}_h$, (c) Ekman upwelling velocity $w_E$, and (d) surface concentration $\tau$, all taken at [$30\degree N, 130\degree W$] in the North Pacific gyre.}
\label{fig:4-way-2yrs}
\end{figure}

We next turn to the physical mechanism leading to the seasonal variation of $\tau$ at the ocean surface. Here we consider three candidate quantities that may be related to the seasonal variation of $\tau$: The first one is the horizontal divergence of the surface water, $\nabla_h \cdot \mathbf{u}_h$, which serves as an indication of flow convergence zone or gyres at the surface. The second one is the Ekman upwelling/downwelling velocity, $w_E$, which can move the particles up and down in the Ekman layer and thus relevant. The Ekman velocity can be computed by taking the curl of surface wind stress (available in ECCOv4r4), which induces a surface motion associated with a vertical flow from continuity \cite{Risien_2008}. The third one is the ocean mixing layer (ML) depth, $h_b$, which affects the vertical mixing of particles. In practice, we compute $h_b$ as the depth where the buoyancy frequency is maximized, i.e., the density gradient is maximized under Boussinesq approximation \cite{Gaspar_1990}.  Figure \ref{fig:4-way-2yrs} shows two-year time series of the three candidate quantities together with $\tau$ at [$30\degree N, 130\degree W$] in the North Pacific gyre, where we see indeed that all three quantities exhibit seasonal variations.  

\begin{figure}[h!]
\centering
    \begin{subfigure}{.6\textwidth}
        \centering
        \includegraphics[width=\linewidth]{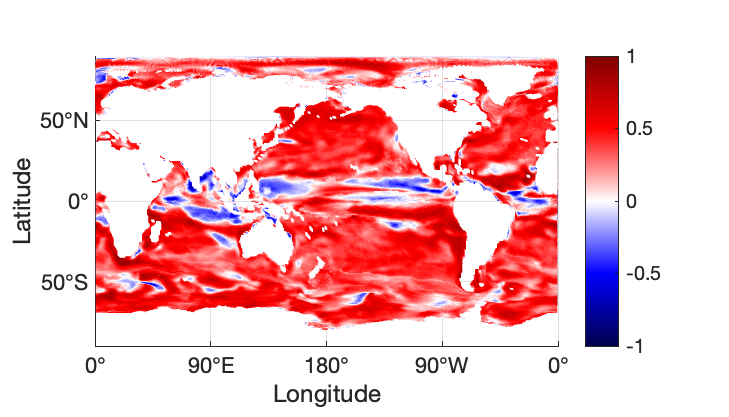}
        \caption{}
        \end{subfigure}
\makebox[\textwidth][c]{%
    \begin{subfigure}{.6\textwidth}
        \includegraphics[width=\linewidth]{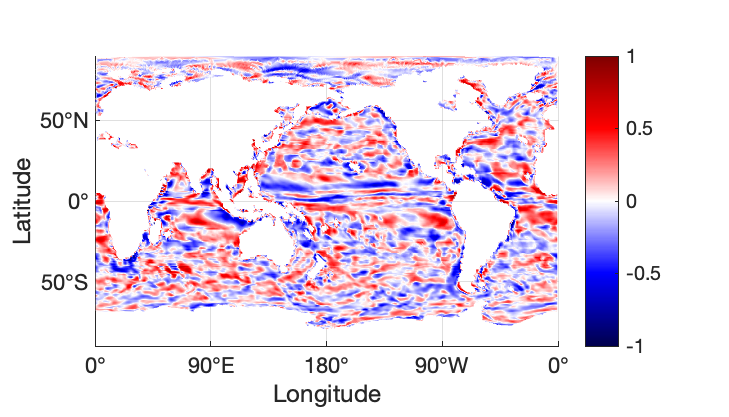}
        \caption{}
        \end{subfigure}%
    \begin{subfigure}{.6\textwidth}
        \includegraphics[width=\linewidth]{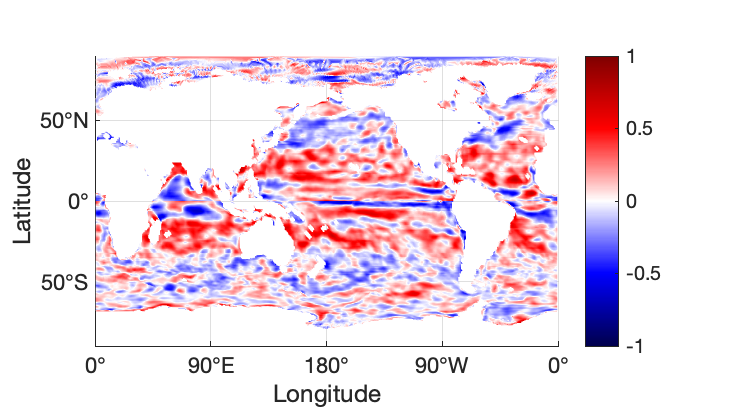}
        \caption{}
        \end{subfigure}
        }
\caption{Maps of correlation coefficients computed over six years (2012-2017) between the surface concentration $\tau$ and (a) the mixed layer depth $h_b$, (b) the horizontal divergence of surface current $\nabla_h \cdot \mathbf{u}_h$, and (c) the Ekman upwelling velocity $w_E$.}
\label{fig:correlations}
\end{figure}

To understand which candidate quantity is most relevant to the seasonal variation of $\tau$, we examine the correlation coefficient between surface $\tau$ and each of the three quantities. We plot in figure \ref{fig:correlations} three global maps of these correlation coefficients, computed for years 2012 through 2017. It is clear that $h_b$ is the most correlated factor, which shows positive correlation coefficient close to 1 in almost all locations in the map. On the other hand, while $\nabla_h \cdot \mathbf{u}_h$ and $w_E$ may be important in forming particle accumulation in gyres, they are not critical for the particle seasonal variation observed. 

\begin{figure}[h!]
\centering
\noindent\includegraphics[width=.5\textwidth]{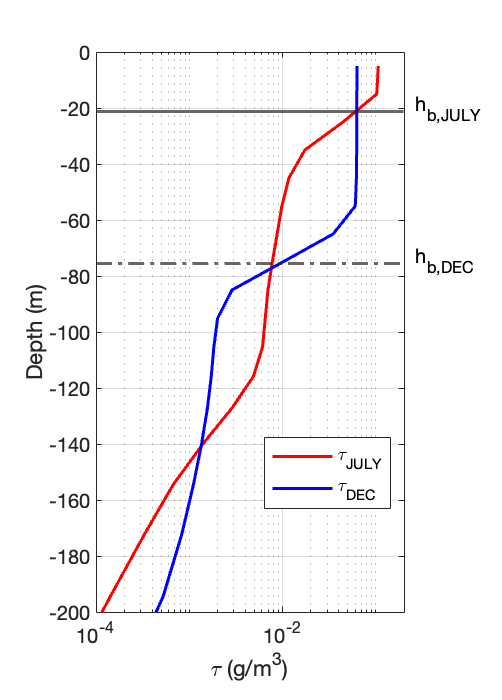}
\caption{Vertical profile of the PE-10 plastic distribution taken at [$30\degree N, 130\degree W$] in July 2017 (red) and December 2017 (blue). The mixed layer depth are indicated in July by solid black line and in December by dashed black line.}
\label{fig:tau_profile}
\end{figure}

The globally favorable correlation between $\tau$ and $h_b$ can be physically explained by the mixing behavior of particles in the ML. Positively buoyant particles such as PE-10 are well mixed in the ML and diminishes rapidly below that as already demonstrated in figure \ref{fig:PE-10_y1k}. It can be expected that as $h_b$ varies seasonally, the particle concentration varies (inversely proportional to $h_b$) conserving the total amount of particles. To confirm this explanation, we plot in figure \ref{fig:tau_profile} the concentration $\tau$ on a vertical slice at [$30\degree N, 130\degree W$] in July (with higher surface concentration) and December (with lower surface concentration). It can be seen that almost uniform concentration is achieved in both July and December with different ML depths, resulting in the different values of $\tau$ at the ocean surface. 

\section{Conclusion and discussion}
\label{sec:conclusion}
We develop a 3D Eulerian model to study microplastic distribution in the global ocean, where the buoyancy effect of the plastic particles is incorporated through a terminal velocity convection term formulated in a mass-conservative way. The terminal velocity depends on both the density and size of the particles, with the latter studied for the first time in this paper. Among the six representative types of particles studied, only positively buoyant particles with sufficient size (e.g., PE with diameter $d=10 \mu m$) accumulate in the five subtropical gyres, forming the garbage patches suggested by earlier works. For all plastic materials studied, as the particle size decreases below a threshold ($\lesssim 1 \mu m$ for PE and PVC, $\lesssim 10 \mu m$ for PP), their distribution converges to that of neutrally buoyant particles, exhibiting weaker accumulation on the surface with a deeper vertical penetration to about 1km. Examination of particle surface concentration also identifies the ocean surface as a filter for particle sizes that behaves differently for different particle densities. In addition, we study the seasonal variation of surface concentration of floating particles (PE-10), which shows reasonable agreement with CYGNSS observation in its covered area in terms of the phase of the seasonal cycle. We further find that this seasonal variation correlates well with the seasonal variation of the ML depth, leading to a physical explanation that the variation of surface concentration is due to a uniform concentration of particles in the ML which changes with the change of ML depth. 

The development of our current model opens a new door for further improvement. Our next goal is to consider simultaneously particles of different densities and sizes, together with models for fragmentation and biofouling. In particular, the fragmentation and biofouling will serve as dissipation mechanisms for surface particle concentration, since the former reduces particle sizes and the latter increases particle density. After these processes are included, we plan to develop a data assimilation interface for the model to consider remote-sensing and trawler measurement data. By tuning parameters in the model (e.g., those related to fragmentation/biofouling, coastal input, etc.), we aim to minimize the model-to-measurement error in a certain form. The final goal is to produce a data-constrained model outputs of microplastic concentration that covers different plastic types and particle sizes.

\section*{Open Research Section}
To reproduce simulations in this paper, one can follow five procedures briefly summarized below.
\begin{enumerate}
   \item Download MITgcm package.
   \item Download ECCOv4r4 dataset (in particular the forcing and initial conditions for MITgcm to reproduce the ECCOv4r4 dataset).
   \item Download additional code (used in step 4) and inputs (used in step 5) at \url{https://github.com/zizien1019/reproduce_eccov4r4_online_68o} to treat additional terms in equation (1). Animations of some simulation results are also provided in the repository.
   \item Compile both the original MITgcm code and additional code in step 3.
   \item Conduct simulations with different inputs on particle properties.
\end{enumerate}

A file uploaded at \url{https://github.com/zizien1019/reproduce_eccov4r4_online_68o/blob/main/Read_me.pdf} contains much more detailed step-by-step instructions.

\bibliographystyle{apacite}
\bibliography{CiteDriveRef.bib}

\acknowledgments
This research is funded in behalf of the Cyclone Global Navigation Satellite System (CYGNSS), National Aeronautics and Space Administration (NASA).

\appendix

\section{Timescale to reach the terminal velocity}
\label{sec:append_t_e_for_w_r}

Consider a round particle moving at an instantaneous velocity $w$, subject to a buoyancy force $B= g (\rho_w-\rho_p) (4\pi/3) (d/2)^3$ and a drag force $D=(12/Re) (\rho_w w^2) \pi (d/2)^2$. Force balance acting on the particle leads to
\begin{linenomath*}
\begin{equation}
 \rho_p \frac{4\pi}{3} (\frac{d}{2})^3 \frac{dw}{dt} = g (\rho_w-\rho_p) \frac{4\pi}{3} (\frac{d}{2})^3 - \frac{12 \mu w}{d} \pi (\frac{d}{2})^2.
\end{equation}
\end{linenomath*}
This forms an ODE for $w$
\begin{linenomath*}
\begin{equation}
 \frac{dw}{dt} = - \frac{18 \mu}{\rho_p d^2} \times w + \frac{g (\rho_w-\rho_p)}{\rho_p}.
 \label{eq:w_dyn}
\end{equation}
\end{linenomath*}
Substituting equation (\ref{eq:wr_stokes}) into (\ref{eq:w_dyn}) gives
\begin{linenomath*}
\begin{equation}
 \frac{dw}{dt} = \frac{18 \mu}{\rho_p d^2} \times (w-w_r).
\end{equation}
\end{linenomath*}
Let $\Delta w(t) = w(t)-w_r$, this yields a solution
\begin{linenomath*}
\begin{equation}
\label{eq:w_r_exponent}
 \Delta w(t) = \Delta w(0) \times \exp{ \{ - \frac{18 \mu}{\rho_p d^2} t \} }.
\end{equation}
\end{linenomath*}
That is, given any difference between the instantaneous velocity and the terminal velocity of a particle, the difference decreases exponentially in time. From the exponent in equation (\ref{eq:w_r_exponent}), the $e$-folding timescale $T_e$ for this velocity difference is
\begin{linenomath*}
\begin{equation}
\label{eq:append_t_e_for_w_r}
{ T_e = \frac{\rho_p d^2}{18 \mu} }.
\end{equation}
\end{linenomath*}
For $\rho_p \sim 1000 kg/m^3$ and $\mu = 0.001 Pa \cdot s$, equation (\ref{eq:append_t_e_for_w_r}) gives $T_e= 0.05 s$ even for large particles of diameter $d=1mm$.

\section{Graphical comparison with earlier studies}
\label{append:graph_compare}
\begin{figure}[h!]
\centering
    \begin{subfigure}{.6\textwidth}
        \centering
        \includegraphics[width=\linewidth]{figs/check_2d_22yrs.png}
        \caption{}
        \end{subfigure}
\makebox[\textwidth][c]{%
    \begin{subfigure}{.6\textwidth}
        \centering
        \includegraphics[width=\linewidth]{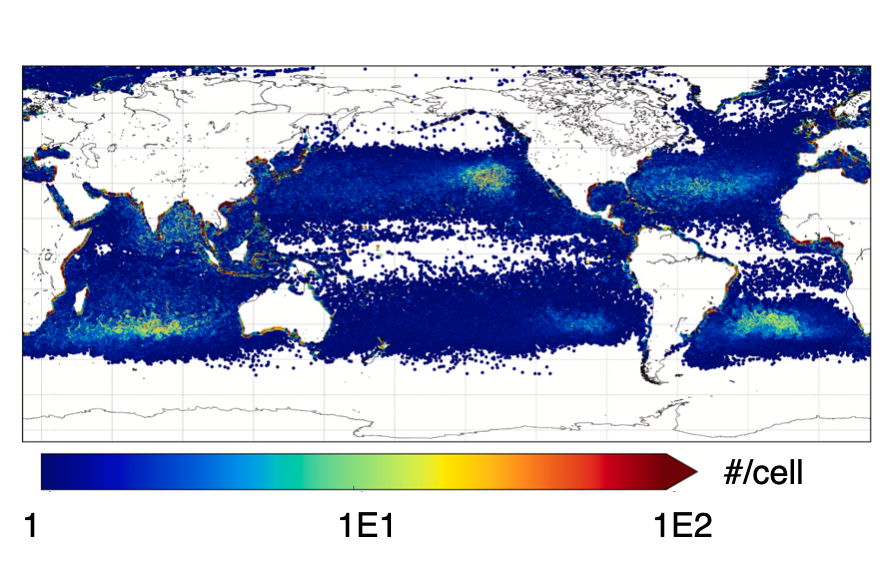}
        \caption{}
        \end{subfigure}%
    \begin{subfigure}{.6\textwidth}
        \centering
        \includegraphics[width=\linewidth]{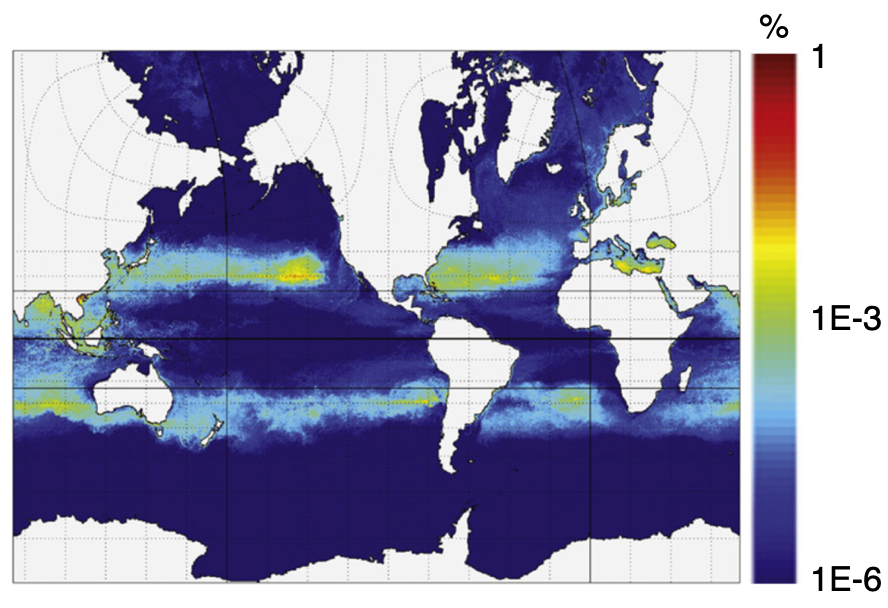}
        \caption{}
        \end{subfigure}
        }
\makebox[\textwidth][c]{%
    \begin{subfigure}{.6\textwidth}
        \centering
        \includegraphics[width=\linewidth]{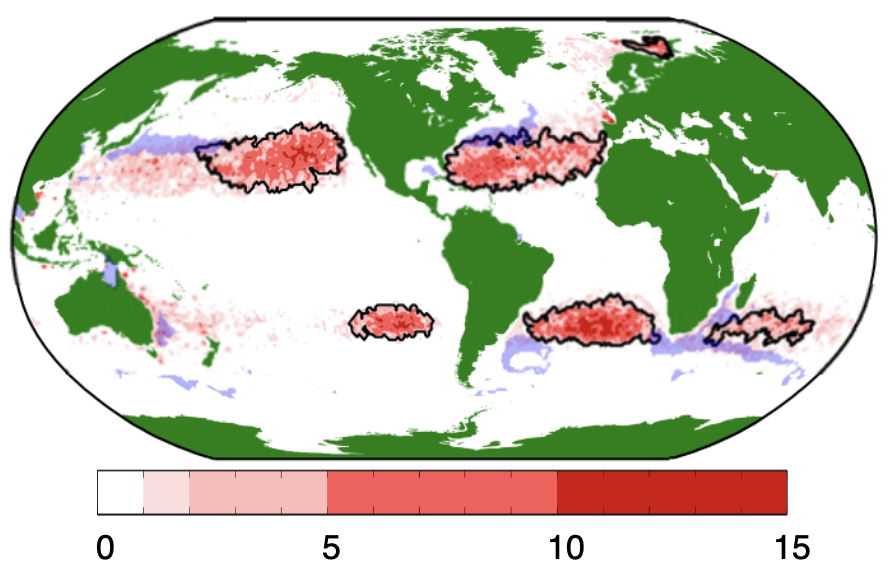}
        \caption{}
        \end{subfigure}%
    \begin{subfigure}{.6\textwidth}
        \centering
        \includegraphics[width=\linewidth]{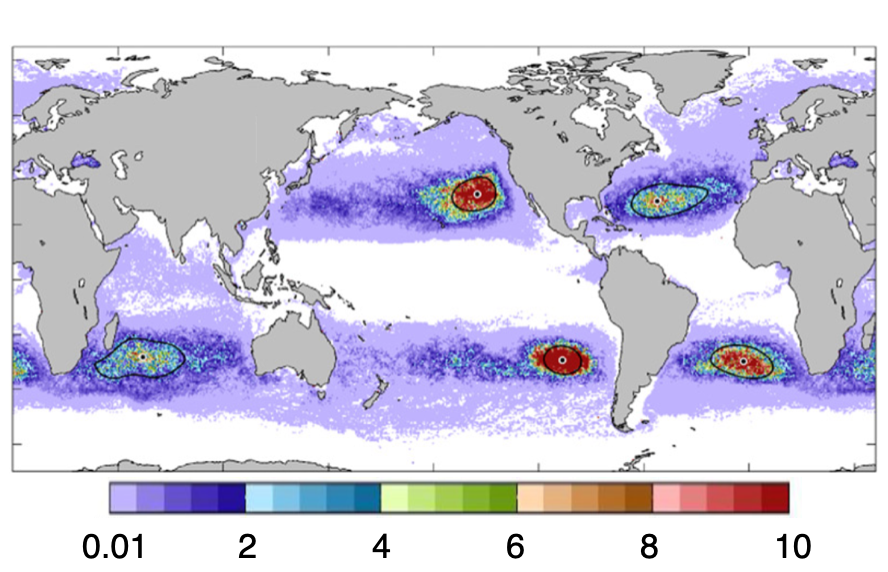}
        \caption{}
        \end{subfigure}
        }
\caption{(a) copy of figure \ref{fig:2D}, (b) particle number density ($\#/[1/12\degree \times 1/12\degree]$ cell) taken from \protect\cite{Chenillat_2021}, (c) particle concentration (unit unspecified) taken from \protect\cite{Lebreton_2012}, (d) tracer amplification factor taken from \protect\cite{van_Sebille_2012}, (e) particle concentration (unit unspecified) obtained by \protect\cite{Maximenko_2012}.}
\label{fig:append_2D_compare}
\end{figure}

\begin{figure}[h!]
\centering
\makebox[\textwidth][c]{%
    \begin{subfigure}{.6\textwidth}
        \centering
        \includegraphics[width=\linewidth]{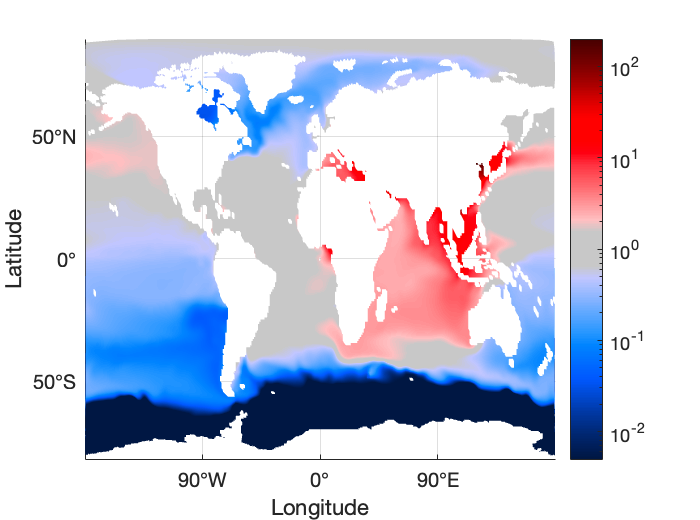}
        \caption{}
        \end{subfigure}%
    \begin{subfigure}{.57\textwidth}
        \centering
        \includegraphics[width=\linewidth]{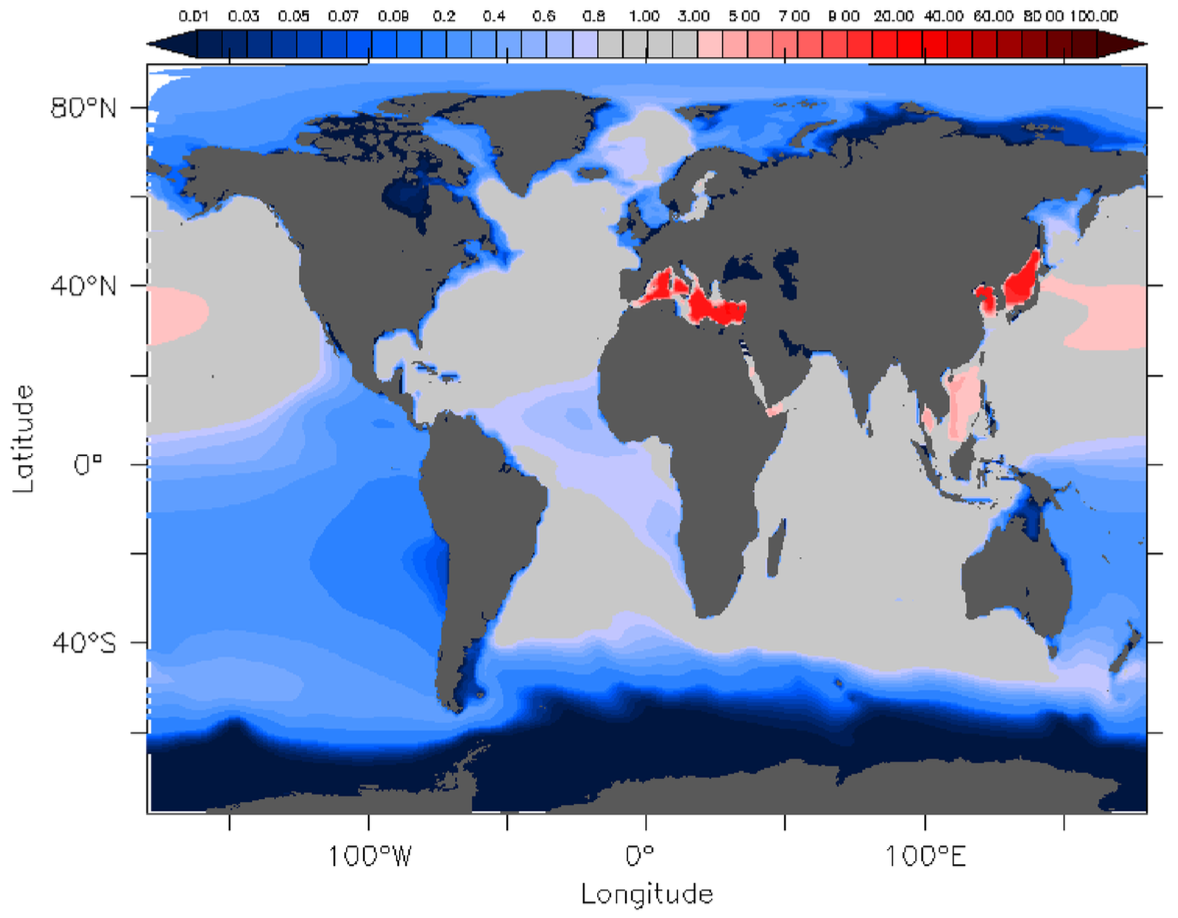}
        \caption{}
        \end{subfigure}
        }
\caption{Neutrally buoyant particle surface distribution, normalized with global mean value, obtained by (a) our model and (b) \protect\cite{Mountford_2019} with the same color bar.}
\label{fig:append_neutral_mountford}
\end{figure}

In this appendix we present comparison of our results in the idealistic cases to earlier relevant studies. Figure \ref{fig:append_2D_compare} shows a comparison of the result in our 2D idealistic case to that in \citeA{Chenillat_2021}, \citeA{Lebreton_2012}, \citeA{Maximenko_2012}, and \citeA{van_Sebille_2012}. Figure \ref{fig:append_neutral_mountford} shows a comparison of the result in our neutrally buoyant case to that in \citeA{Mountford_2019} with the same color bar. 

%
%


%
%
%
%
%

\end{document}